\documentclass[review]{elsarticle}

\usepackage{hyperref}

\usepackage{comment}
\usepackage{graphicx}
\usepackage{xcolor}
\usepackage{amsmath}
\usepackage{mathtools}
\usepackage{textcomp}
\usepackage{amsfonts}
\usepackage{microtype}
\usepackage{multirow}
\usepackage[T1]{fontenc} 
\usepackage{tabularx,ragged2e,booktabs,babel}
\usepackage{array}
\usepackage{verbatim}
\usepackage{natbib}
\usepackage{csquotes}
\usepackage{tablefootnote}
\usepackage{siunitx}
\usepackage{csvsimple}
\usepackage{booktabs}
\usepackage{subfig}

\journal{Journal of \LaTeX\ Templates}

\usepackage{numcompress}

\newcommand{\wfig}[1]{Fig.~\ref{fig:#1}}
\newcommand{\wtab}[1]{Table \ref{tab:#1}}

\begin{document}

\begin{frontmatter}

\title{Automated Detection of Abnormalities from an EEG Recording of Epilepsy Patients With a Compact Convolutional Neural Network}
\tnotetext[mytitlenote]{Fully documented templates are available in the elsarticle package on \href{http://www.ctan.org/tex-archive/macros/latex/contrib/elsarticle}{CTAN}.}

\author[tuat]{Taku Shoji}
\author[nerima]{Noboru Yoshida}
\author[tuat]{Toshihisa Tanaka\corref{mycorrespondingauthor}}
\cortext[mycorrespondingauthor]{Corresponding author}
\ead{tanakat@cc.tuat.ac.jp}


\address[tuat]{Tokyo University of Agriculture and Technology, Tokyo 184-8588, Japan}
\address[nerima]{Department of Pediatrics, Juntendo University Nerima Hospital, Tokyo, 177-8521 Japan}

\begin{abstract}
Electroencephalography (EEG) is essential for the diagnosis of epilepsy, but it requires expertise and experience to identify abnormalities. It is thus crucial to develop automated models for the detection of abnormalities in EEGs related to epilepsy. This paper describes the development of a novel class of compact convolutional neural networks (CNNs) for detecting abnormal patterns and electrodes in EEGs for epilepsy. The designed model is inspired by a CNN developed for brain-computer interfacing called multichannel EEGNet (mEEGNet). Unlike the EEGNet, the proposed model, mEEGNet, has the same number of electrode inputs and outputs to detect abnormal patterns. The mEEGNet was evaluated with a clinical dataset consisting of 29 cases of juvenile and childhood absence epilepsy labeled by a clinical expert. The labels were given to paroxysmal discharges visually observed in both ictal (seizure) and interictal (nonseizure) durations. Results showed that the mEEGNet detected abnormalities with the area under the curve, F1-values, and sensitivity equivalent to or higher than those of existing CNNs. Moreover, the number of parameters is much smaller than other CNN models. To our knowledge, the dataset of absence epilepsy validated with machine learning through this research is the largest in the literature.
\end{abstract}


\end{frontmatter}

\section{Introduction}

Epilepsy is a chronic neurological disorder that affects approximately 50 million people worldwide \citep{WHO}.
An electroencephalogram (EEG) is a crucial tool in the diagnosis of epilepsy.
Depending on its invasiveness, EEG can be classified into either scalp EEG or intracranial EEG (iEEG).
A scalp EEG is performed using electrodes attached to the scalp and is used for an early epilepsy diagnosis. On the other hand, iEEG is performed through placing electrodes inside the skull by craniotomy.
Intracranial EEG is used to identify the primary site of epilepsy in patients whose seizures are not controlled by medication and who may benefit from its surgical removal of the primary site of epilepsy \citep{Parvizi2018}.


To diagnose epilepsy based on EEG measurements, a trained specialist (epileptologist) must visually read the EEG and distinguish between normal and abnormal epileptic waveforms. This process is highly specialized, time-consuming, and laborious and is a significant burden for epileptologists. Therefore, there is a need to detect epileptic intervals in EEGs automatically. In the past decades, machine learning's effectiveness has been widely recognized because it can automatically learn the features necessary for detection by collecting a sufficient amount of data. In the context of machine learning-based approaches, most of the automated techniques for scalp EEG are designed for seizure or seizure onset detection \citep{Siddiqui2020} and waveforms (such as spike) detection \citep{AbdEl-Samie2018}. In particular, methods based on convolutional neural networks (CNNs), one of the deep neural networks (DNNs), have shown high performance in detection tasks \citep{Hossain2019, Acharya2018, Segundo2019, Zhou2018, Fukumori2019, Johansen2016}.


In terms of the diagnosis of epilepsy, ictal (seizure) EEGs are crucial, but interictal (nonseizure) EEGs are also essential. Abnormal waveforms are also observed in a patient's EEG during interictal states. Even though these abnormal patterns are the basis of diagnosis, they are intermittent and infrequent in an entire EEG recording, typically lasting about half an hour. Looking for abnormal patterns from an entire EEG is time-consuming for epileptologists. However, there are few studies on the automatic detection of abnormal patterns in EEGs of epilepsy patients.
\citet{Sakai2020} proposed a CNN model to detect abnormalities in EEGs of child absence epilepsy patients. CNNs typically have many parameters. They require a large labeled dataset for training, although, similar to other medical data, it is difficult to collect a significant number of epileptic EEGs. This points to a need to develop a compact CNN model that is suitable for EEG. 

Recently, in the field of brain-computer interfacing, a CNN model for multichannel EEG called EEGNet, was successfully proposed \citep{Lawhern2018}. The EEGNet demonstrated its efficiency in the classification of EEG to different mental states such as motor imagery. This paper proposes using a novel class of the EEGNet called multichannel EEGNet (mEEGNet), a compact CNN model based on EEGNet, which can detect EEG abnormalities in epilepsy for each electrode. To validate the mEEGNet, we experimented with a practical dataset consisting of labeled EEGs from 29 cases (19 patients) of childhood absence epilepsy (CAE) and juvenile absence epilepsy (JAE), both of which are major childhood epilepsy syndromes. The performance is compared with conventional models using actual EEG data from epilepsy patients to show its effectiveness in terms of F1-score, sensitivity, specificity, and AUC.

Our contributions can be summarized as follows. We have revealed that a compact CNN model based on the EEGNet can identify abnormal patterns in EEG of CAE and JAE patients with higher accuracies than other complicated architectures. We tested several machine learning models with an original dataset consisting of 19 epilepsy patients in the validation. To our knowledge, the dataset used in the validation is the largest in the literature. This technology may enable epileptologists to extract ``keyframes'' from EEG recordings up to 40 minutes, in which abnormal patterns appear infrequent.

\section{Related Works}



In engineering, seizure detection from the EEGs of epilepsy patients is an essential topic in epileptic signal analysis, and most published works have used public datasets such as the Bonn dataset \cite{Andrzejak2001} and the CHB-MIT corpus \cite{Goldberger2000}. In particular, the Bonn dataset consists of 100 single-channel EEG segments of 23.6 s duration. For each segment, the label of normal, interictal, or ictal EEG is given. This is the easy go-to for computer simulations; therefore, hundreds of papers have been published in the engineering context\footnote{For recent works, see \citep{Sharma2020,Siddiqui2020,Shoeibi2021}, for example.}. Although the Bonn dataset contributed to the development of classification algorithms, the data format is incomplete and far from a practical clinical format; thus, it is difficult to use in clinical settings.

On the other hand, the CHB-MIT corpus contains multi-electrode EEGs of 23 subjects with labels of seizure duration. This corpus contributed significantly to the development of neural network algorithms for seizure detection in EEGs using CNNs and recurrent neural networks \citep{Hossain2019, Thodoroff2016, Acharya2018, Segundo2019, Zhou2018}. However, the label is only applied to ictal (seizure) times. For diagnosis of epilepsy, abnormal waveforms (such as spike waves) during the interictal state are also essential. To detect abnormalities, some of the authors of the current paper have proposed a two-dimensional CNN (ScalpNet) with convolutional kernels in both time and electrode directions for detecting abnormal epileptic segments \citep{Sakai2020}.

In terms of waveform detection, epileptic spike detection is a major problem. Spike waves appear in the non-paroxysmal scalp EEGs of epileptic patients and are essential biomarkers in diagnosis. Their automatic detection could reduce the burden on specialists \citep{Johansen2016,Fukumori2019,Maurice2020}.
\citet{Johansen2016} used CNN to classify EEGs preprocessed with high-pass and notch filters to detect spike waves.
\citet{Fukumori2019} used CNN to classify abnormal epileptic spikes and artifacts.

Abnormality can be labeled for a whole recording of EEG. Some works used a dataset where every EEG recording had one label: normal or abnormal. Van Leeuwen et al.~\cite{VanLeeuwen2019} used their own clinical dataset consisting of 15-min EEG recordings labeled normal or abnormal. Every 1-min segment from a 15-min EEG was determined normal or abnormal by CNN, and the segment-wise features were aggregated to make a final decision for the input EEG. Yıldırım et al.~\cite{Yldrm2020} used TUH abnormal corpus to test their CNN that needs the first 1-min of the EEG recording to determine normal or abnormal. Both methods used balanced datasets where the numbers of normal and abnormal labels are similar.

Our aim is different from the above mentioned studies. We detect abnormal patterns from an entire EEG recordings of absence epilepsy patients, because epileptologists often visually review a long-term EEG to find abnormal patterns (epileptiform discharges and slow waves) for the diagnosis. However, the abnormal patterns in EEGs are intermittent, so most intervals are normal.

\section{Methods}
\subsection{EEG Data}


In this study, scalp EEG was measured in 19 patients with epilepsy at Juntendo University Nerima Hospital, Japan. Nine of the 19 patients were diagnosed with JAE and 10 with CAE. The dataset consisted of 29 cases, including multiple measurements from the same patient on different occasions.
The sex, age, and disease name are shown in \wtab{dataset}. The EEG No. in \wtab{dataset} is a serial number, and the alphabetical letters were used as patient identifiers, as the dataset includes multiple measurements from the same patient.

For all 29 cases shown in \wtab{dataset}, the sampling frequency was 500 Hz, and the number of electrodes was 16. A clinical expert visually inspected all cases to find paroxysmal discharges \citep{Panayiotopoulos1999,Mariani2011} and labeled the time duration and electrodes of abnormalities. These paroxysmal discharges were observed in both ictal and interictal EEGs, and were sometimes induced by hyperventilation \citep{Kessler2019}.
The EEG recording and analysis were approved by the Ethics Committee of Tokyo University of Agriculture and Technology and the Ethics Committee of Juntendo University Nerima Hospital in Tokyo, Japan. Written informed consent was obtained from the patients and caretakers.

\begin{table}[t]
  \centering
  \caption{
  Details of the dataset: The column of EEG No.\ consists of the serial number of this dataset and the patient's unique alphabet. Different numbers with the same alphabet indicate that the EEGs were performed on different days on the same patient. CAE and JAE indicate childhood absence epilepsy and juvenile absence epilepsy, respectively. For each case, the time duration labelled as abnormal by a clinical expert is denoted by $T_a$ and the total recorded time is denoted by $T$ in seconds.
}
  \label{tab:dataset}

      \begin{tabular}{ccccrrS} \hline
      \begin{tabular}{l} EEG \\ No. \end{tabular} & Sex & \begin{tabular}{c} Age \\ {}[yrs] \end{tabular} & Type & \begin{tabular}{c@{\,}} Abnormal \\ $T_a$ [s] \end{tabular} & \begin{tabular}{c@{\,}} Recording \\ $T$ [s] \end{tabular} & \begin{tabular}{c@{\,}} Ratio \\ $T_a/T$ \end{tabular} \\ \hline
      1A  & M & 6  & CAE & 32.1 & 1,976 & 1.62\% \\  
      2A  & M & 7  & CAE & 0 & 1,754 & 0\% \\  
      3B  & F & 7  & JAE & 53.8 & 1,873 & 2.87\% \\  
      4C  & F & 8  & CAE & 49.8 & 1,569 & 3.17\% \\  
      5D  & M & 8  & CAE & 7.48 & 1,749 & 0.428\% \\  
      6D  & M & 8  & CAE & 38.5 & 1,887 & 2.04\% \\  
      7D  & M & 8  & CAE & 8.53 & 483 & 1.77\% \\  
      8E  & F & 10 & JAE & 164 & 2,509 &  6.54\% \\  
      9E  & F & 10 & JAE & 21.7 & 1,793 & 1.21\% \\  
      10F & M & 9  & JAE & 10.5 & 737 & 1.42\% \\  
      11B & F & 12 & JAE & 8.60 & 1,770 & 0.486\% \\  
      12F & M & 9  & JAE & 0 & 1,803 & 0\% \\  
      13G & F & 14 & JAE & 4.74 & 1,850 & 0.256\% \\  
      14H & F & 9  & JAE & 133 & 1,787 & 7.44\% \\  
      15I & F & 12 & JAE & 33.4 & 1,779 & 1.88\% \\  
      16G & F & 15 & JAE & 9.05 & 1,726 & 0.524\% \\  
      17J & F & 8  & JAE & 123 & 1,844 & 6.67\% \\  
      18K & M & 12 & JAE & 55.0 & 1,515 & 3.63\% \\  
      19L & F & 8  & JAE & 95.5 & 1,690 & 5.65\% \\  
      20L & F & 8  & JAE & 35.8 & 1,744 & 2.05\% \\  
      21M & F & 7  & CAE & 122 & 1,296 & 9.41\% \\  
      22M & F & 7  & CAE & 3.42 & 1,950 & 0.175\% \\  
      23N & F & 11 & CAE & 19.5 & 438  & 4.45\% \\  
      24O & F & 6  & CAE & 59.3 & 1,670 & 3.55\% \\  
      25P & F & 9  & CAE & 46.0 & 1,366 & 3.37\% \\  
      26Q & F & 9  & CAE & 34.0 & 1,648 & 2.06\% \\  
      27R & M & 6  & CAE & 30.7 & 1,644 & 1.87\% \\  
      28S & M & 8  & CAE & 8.40 & 483 & 1.74\% \\  
      29A & M & 6  & CAE & 34.6 & 1,644 & 2.10\% \\  
      \hline
      Total & & & & 1,242.4 & 45,977 & 2.70\% \\ \hline
      \end{tabular}
\end{table}

\subsection{mEEGNet: A Model for Detecting Epileptic Abnormalities in EEG}

\begin{figure}
    \centering
    \includegraphics[width=\columnwidth,clip]{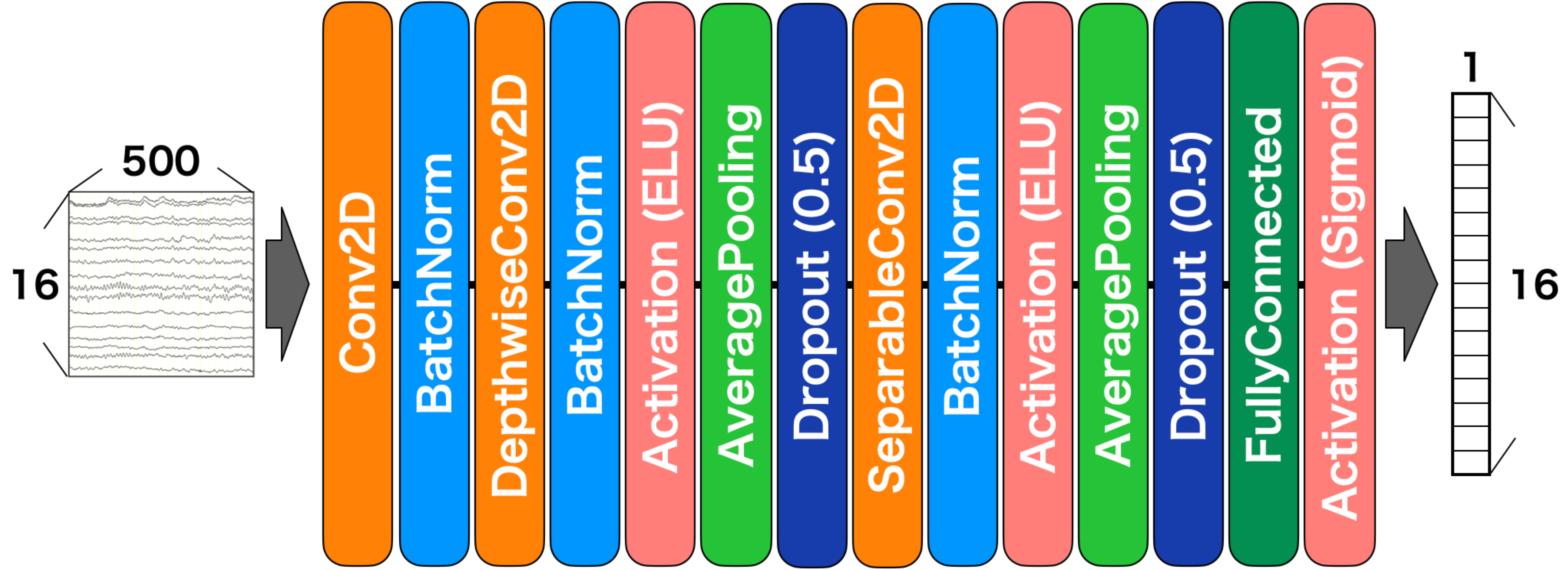}
    \caption{The pipeline of the proposed mEEGNet. Details for each layer are given in Table \ref{tab:mEEGNet}.}
    \label{fig:meegnet}
\end{figure}

This paper proposes a CNN architecture called mEEGNet for detecting epileptic abnormal EEGs. mEEGNet is an architecture inspired by EEGNet \citep{Lawhern2018}, which was originally proposed for brain-computer interfaces. The pipeline of mEEGNet is illustrated in Fig.~\ref{fig:meegnet}, and the details of mEEGNet are listed in Table \ref{tab:mEEGNet}. The proposed mEEGNet has an input layer with a size of $16 \times 500$, which adapts to 16 electrodes (international 10-20 system of the electrode location) and a one-second time window. Also, the output layer has 16 units. This implies that the mEEGNet makes a positive or negative decision every second at each electrode. The mEEGNet consists of three convolutional layers and one fully connected layer.

\begin{table}[t]
\caption{Architecture of mEEGNet. The effect of the kernel size of Conv2D (1, 250) is evaluated in the experiment (See Fig.~\ref{fig:kernel_size}).}
\centering
\begin{tabular}{l|rr} \hline
Layer & \begin{tabular}[l]{@{}l@{}} \# filters / \\ Kernel size \end{tabular} & Output shape\\ \hline
Input & & (1, 16, 500) \\
\textbf{Conv2D} & 8 / (1, \textbf{250}) &  (8, 16, 500) \\
BatchNorm & & (8, 16, 500) \\
DepthwiseConv2D & 16 / (16, 1) & (16, 1, 500) \\
BatchNorm &  & (16, 1, 500)  \\
Activation (ELU) & & (16, 1, 500) \\
AveragePooling2D & & (16, 1, 125) \\
Dropout & & (16, 1, 125) \\
SeparableConv2D & 16 / (1, 16) & (16, 1, 125) \\
BatchNorm &  & (16, 1, 125) \\
Activation (ELU) & & (16, 1, 125) \\
AveragePooling2D & & (16, 1, 15) \\
Dropout & & (16, 1, 15) \\
Dense & & 16 \\
Activation (Sigmoid) & & 16 \\
\hline
\end{tabular}
\label{tab:mEEGNet}
\end{table}

The first Conv2D layer of the mEEGNet is a convolutional layer with a kernel only in the time direction. The purpose of the first layer is to extract features in the frequency domain of an EEG. Since the kernel size in the time direction is set to 1/2 of the sampling frequency, it can extract frequency features above 2 Hz. The kernel size will be further explored in the simulation, as described later in this paper.
The second convolutional layer, DepthwiseConv2D, is a convolutional layer with kernels only in the direction of the electrodes. It aims to learn the relationship between electrodes by weighting and adding the features of each electrode obtained in the first convolutional layer.
The third convolutional layer is SeparableConv2D.
Separable convolution is a convolutional layer that combines depthwise convolution and pointwise convolution. A dropout layer follows to suppress overfitting.


The significant difference from EEGNet \citep{Lawhern2018} is the second stage following SeperableConv2D.
There are an equal number of corresponding electrodes in the inputs and outputs of mEEGNet. The output uses a sigmoid function for the activation function of the final layer to indicate electrode abnormalities.

\subsection{Weighted Loss Functions}
\label{s:imbalance}

The binary cross-entropy (BCE) is a widely used loss function for training neural networks with the binary output. The loss function of BCE is given as
\begin{equation}
    CE(p_t) = -\log p_t
\end{equation}
where $p_t$ describes the model's estimated probability (denoted by $p$) for one class and $1-p$ for the other class. 
One of the significant challenges in anomaly detection is the imbalance between normal labels and abnormal labels.
When a dataset that does not have an equal number of data per label is used to train a classifier (detector), it tends to result in an unbalanced classifier that has a very high detection rate for the majority class and a low detection rate for the minority class \citep{He2009}. To solve this problem, learning with weighted loss functions such as focal loss (FL) \citep{Lin2020} and class-balanced loss (CBF) \citep{Cui2019} has been proposed.
This paper uses not only binary cross-entropy (BCE) but also FL defined as
\begin{equation}
    FL(p_t) = -\alpha ( 1-p_t)^\gamma \log p_t
\end{equation}
and CBF defined as
\begin{equation}
    CBF(p_t) = -\frac{1-\beta}{1-\beta^n} ( 1-p_t)^\gamma \log p_t
\end{equation}
to check their impact on the detection performance of mEEGNet. In the above loss functions, $\alpha$, $\gamma$, and $\beta$ are hyperparameters, and $n$ is the number of training samples.

\subsection{Comparison}

We conducted simulations to compare the evaluation criteria (AUC, F1-value, sensitivity, and specificity) for the following methods:
\begin{itemize}
    \item mEEGNet (proposed);
    \item ScalpNet~ \citep{Sakai2020};
    \item Zhou et al.'s model~\citep{Zhou2018};
    \item Hossain et al.'s model~\citep{Hossain2019};
    \item SVM with Gaussian kernel~\citep{Cristianini2008}.
\end{itemize}

\subsubsection{ScalpNet}

\begin{figure}
    \centering
    \includegraphics[width=\textwidth,clip]{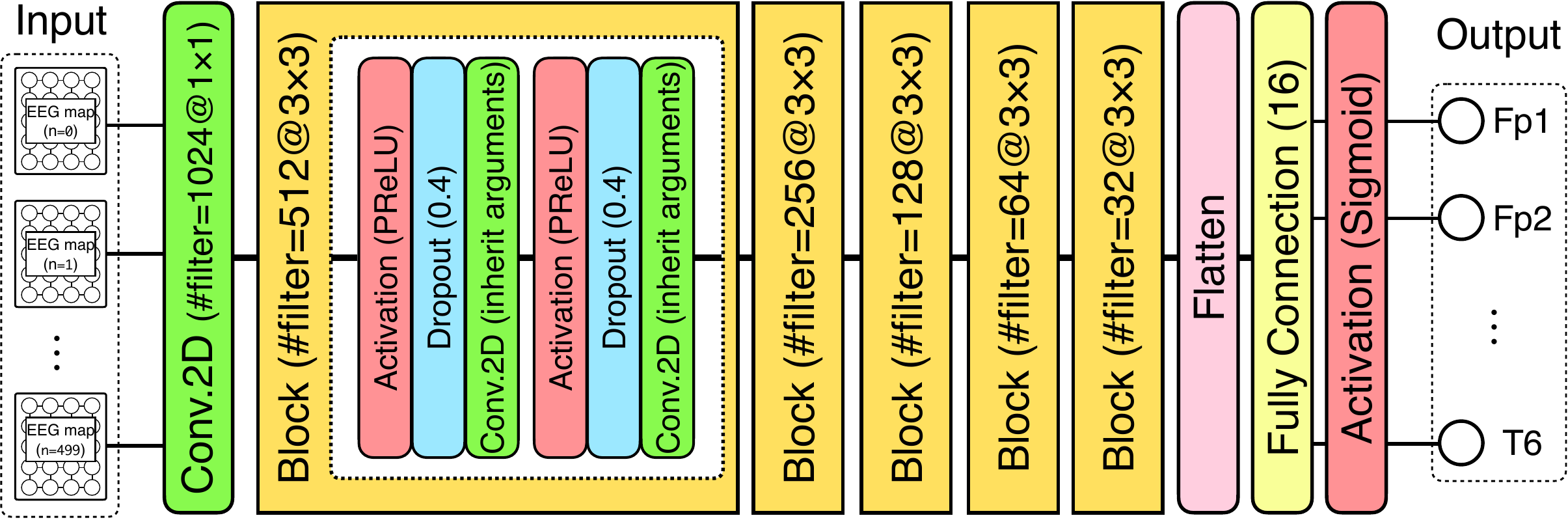}
    \caption{The architecture of ScalpNet \citep{Sakai2020}}
    \label{fig:scalpnet}
\end{figure}

ScalpNet is a simple CNN architecture proposed by \citet{Sakai2020}, and the network structure is illustrated in Fig.~\ref{fig:scalpnet}. The main feature of this network is that the 16 electrodes of the international 10-20 system is rearranged as a two-dimensional array of $5 \times 4$ by assuming that the distance between the electrodes in the 16 electrodes was approximately equal.

\subsubsection{Other CNN-based Models}
Note that Hossain et al.'s CNN is a two-class classification model for seizures and non-seizures, while Zhou et al.'s CNN is a three-class classification model for ictal, pre-ictal, and non-ictal signals. The input samples' size in the time direction is 512 dimensions in Hossain et al.'s CNN and 256 dimensions in Zhou et al.'s CNN, which is different from mEEGNet. Therefore, to compare the performance of Zhou et al.'s model with that of mEEGNet, we constructed a CNN in which the input size, output size, and activation function of Zhou et al.'s model and Hossain et al.'s model are aligned with those of mEEGNet. In Zhou et al.'s model, the input signal can be in the time domain (Zhou-time) or the frequency domain (Zhou-freq).

\subsubsection{Support Vector Machine}
A support vector machine (SVM) is a well-known standard in machine learning that utilizes a kernel function to map a feature space to a non-linear high dimensional space.
We used the Gaussian function as a kernel function. The parameters (regularization parameter, $C$, and kernel coefficient, $\gamma$) were determined through the grid search in the parameter set defined as
$C=[0.1,1,5,10,50,100,500,1000]$ and $\gamma =[0.01,0.1,1,5,10,100]$ at every validation (5-fold cross-validation and leave-one-case-out validation, as described in the following).







\subsection{Evaluation}
\label{s:metrics}


The implementations of neural network models and SVM were written in Keras \cite{chollet2015keras} with a Tensorflow backend \cite{abadi2016tensorflow} and in cuML \cite{raschka2020machine}, respectively. We ran our experiments with Amazon Linux 2 on Amazon Web Service (AWS) EC2 P3 instances.

This paper examines the model's performance through two types of validation: 5-fold cross-validation and leave-one-case-out validation. In the 5-fold cross-validation, all the EEGs from the 29 cases shown in \wtab{dataset} were randomly divided into five blocks; one of the five blocks was test data, and the remaining four blocks were training data. Thus, the cross-validation was conducted five times independently with randomized divisions for 25 training and testing sessions in total. In the leave-one-case-out validation, one of the 29 cases shown in \wtab{dataset} was the test data and the remaining 28 cases were the training data. The evaluation criteria in both validations were the area under the curve (AUC), the F1-value, sensitivity, and specificity.

In the both validation methods, the hyperparameters of the loss functions (FL and CBF) were grid-searched in terms of F1-values through 3-fold cross-validation in the training data. For FL, the candidates in the grid search were $\alpha = \{ 0.25, 0.333, 0.5, 0.666, 0.75\}$ and $\gamma = \{ 0, 0.1, 0.3, 0.5, 1.0, 1.5, 2.0 \}$, and for CBF, the candidates were $\beta = \{ 0.9, 0.99, 0.999, 0.9999 \}$ and $\gamma = \{ 0, 0.1, 0.3, 0.5, 1.0, 1.5, 2.0 \}$.


As shown in Table \ref{tab:mEEGNet}, the first layer of mEEGNet (Conv2D) has 250 parameters, which is the largest among the layers. Thus, a reduction of the number of parameters in this layer could contribute to simplification of the mEEGNet. We evaluated the performance of the mEEGNet with kernel sizes of 10, 50, 125, and 250 in the first layer.
For the AUC and the F1-value, the Friedman test \citep{Friedman1937} was performed with the significance level set at $p<0.05$. The Nemenyi test \citep{Nemenyi1963} was performed as a post hoc test, and combinations of models with significant differences were also investigated.

\section{Results}

\begin{table}
\centering
\caption{Results of the 5-fold cross validation method.\label{tab:5fold}}
\begin{tabular}[t]{c|c|cccc} \hline
\multicolumn{2}{c|}{Model} & AUC  & F1  & Sensitivity  & Specificity  \\ \hline
\multirow{3}{*}{mEEGNet} & BCE & $\textbf{.990} \pm .00386$  &  $\textbf{.837} \pm .0155$  &  $\textbf{.824} \pm .0254$  &  .996 $\pm$ .000631  \\ \cline{2-6}
& Focal & $.989 \pm .00304$  &  $.828 \pm .0192$  &  $.809 \pm .0414$  &  996 $\pm$ .00163  \\ \cline{2-6}
& CBFocal & $.989 \pm .00302$  &  $.833 \pm .0152$  &  .819 $\pm$  .0251  &  .996 $\pm$ .00107  \\ \hline
\multirow{3}{*}{\begin{tabular}{c}
ScalpNet \\ \citep{Sakai2020} \end{tabular}} & BCE &  $.969 \pm .0111$  &  $.824 \pm .0170$  &  .823 $\pm$ .403  &  .995 $\pm$ .00228  \\ \cline{2-6}
& Focal & $.894 \pm .176$  &  $.683 \pm .305$  &  .717 $\pm$ .323  &  .994 $\pm$ .00408  \\ \cline{2-6}
& CBFocal & $.858 \pm .202$  &  $.626 \pm .360$  &  .625 $\pm$ .361  &  .996 $\pm$ .00241  \\ \hline
\multicolumn{2}{c|}{\citet{Hossain2019}} & $.838 \pm .158$  &  $.453 \pm .378$  &  .412 $\pm$ .344  &  .998 $\pm$ .00185  \\ \hline
\multirow{2}{*}{\citet{Zhou2018}} & Time & $.590 \pm .0183$  &  $0 \pm 0$  &  0 $\pm$ 0  &  \textbf{1.00} $\pm$ 0.000  \\ \cline{2-6}
& Freq. & $.974 \pm .00563$ &  $.798 \pm .0202$ &  .770 $\pm$ .0379 &  .996 $\pm$ .000790 \\ \hline
\multicolumn{2}{c|}{SVM} &  $.921 \pm .00322$  &  $.833 \pm .00811$  &  .308 $\pm$ .00698  &  .999 $\pm$ .0000950  \\ \hline
\end{tabular}
\end{table}

\begin{table}
\centering
\caption{Results of the leave-one-case-out method.\label{tab:loco}}
\begin{tabular}[t]{c|c|cccc} \hline
\multicolumn{2}{c|}{Model} & AUC  & F1  & Sensitivity  & Specificity  \\ \hline
\multirow{3}{*}{mEEGNet} & BCE &  $\textbf{.976} \pm .0739$  &  \textbf{.760} $\pm$ .271  & \textbf{.773} $\pm$ .275  & .995 $\pm$ .00768 \\ \cline{2-6}
& Focal &  .964 $\pm$ .0935  &  .733 $\pm$ .290  & .733 $\pm$ .296  & .996 $\pm$ .00725 \\ \cline{2-6}
& CBFocal & .965 $\pm$ .0922  &  .734 $\pm$ .296  & .744 $\pm$ .302 & .995 $\pm$ .00677 \\ \hline
\multirow{3}{*}{\begin{tabular}{c}
ScalpNet \\ \citep{Sakai2020} \end{tabular}} & BCE &  .894 $\pm$ .266  &  .684 $\pm$ .336  &  .709 $\pm$ .348 & .995 $\pm$ .00877  \\ \cline{2-6}
& Focal &  .858 $\pm$ .284  &  .645 $\pm$ .376  & .691 $\pm$ .386  &  .994 $\pm$ .00949   \\ \cline{2-6}
& CBFocal & .858 $\pm$ .213  &  .557 $\pm$ .406  & .591 $\pm$ .410  & .995 $\pm$ .00891 \\ \hline
\multicolumn{2}{c|}{\citet{Hossain2019}} & .838 $\pm$ .227  &  .508 $\pm$ .370  & .501 $\pm$ .374  & .996 $\pm$ .00633   \\ \hline
\multirow{2}{*}{\citet{Zhou2018}} & Time & .587 $\pm$ .103  &  0 $\pm$ 0  & 0 $\pm$ 0  & \textbf{1.00} $\pm$ 0.000 \\ \cline{2-6}
& Freq. &  .975 $\pm$ .0686  &  .733 $\pm$ .285 & .757 $\pm$ .290 & .995 $\pm$ .00617 \\ \hline
\multicolumn{2}{c|}{SVM} &  .910 $\pm$ .109  &  .667 $\pm$ .249  & .616 $\pm$ .240  &   .996 $\pm$ .00465 \\ \hline
\end{tabular}
\end{table}

\subsection{AUC, F1-value, Sensitivity, and Specificity}
Tables \ref{tab:5fold} and \ref{tab:loco} list the results of averaged scores of the 5-fold cross validation and the leave-one-case-out methods, respectively. In Table \ref{tab:5fold}, all the values are the ground average with the standard deviation across the cross validation repeated five times ($N=25$).
Table \ref{tab:loco} shows the ground average ($N=27$) with the standard deviation over all but two cases: 2A and 12F, for which all the models returned normal (true positives and false negatives were both zero).

In both validation methods, the proposed mEEGNet with the loss function of BCE achieved the highest AUCs of $0.990$ (5-fold) and $0.976$ (leave-one-case-out) and the highest F1 scores of $0.837$ (5-fold) and $0.760$ (leave-one-case-out) among all compared models.
In medical evaluations, sensitivity and specificity are crucial measures in classification (diagnosis). Tables \ref{tab:5fold} and \ref{tab:loco} exhibit very high specificity values of nearly one. This implies that all the models work well in detecting normal EEG states. Thus, the most important measure to evaluate performance is sensitivity. In the both validation methods, the proposed mEEGNet with BCE provided the highest scores.
From the results, Hossain et al.'s model, Zhou et al's model with time, and SVM are inappropriate for the detection of abnormalities. Moreover, in terms of the loss function, BCE is the best among the three functions.

\subsection{Kernel Size of the First Layer}
\label{s:result_ks}

\begin{figure}[t]
  \centering
  \includegraphics[width=\textwidth]{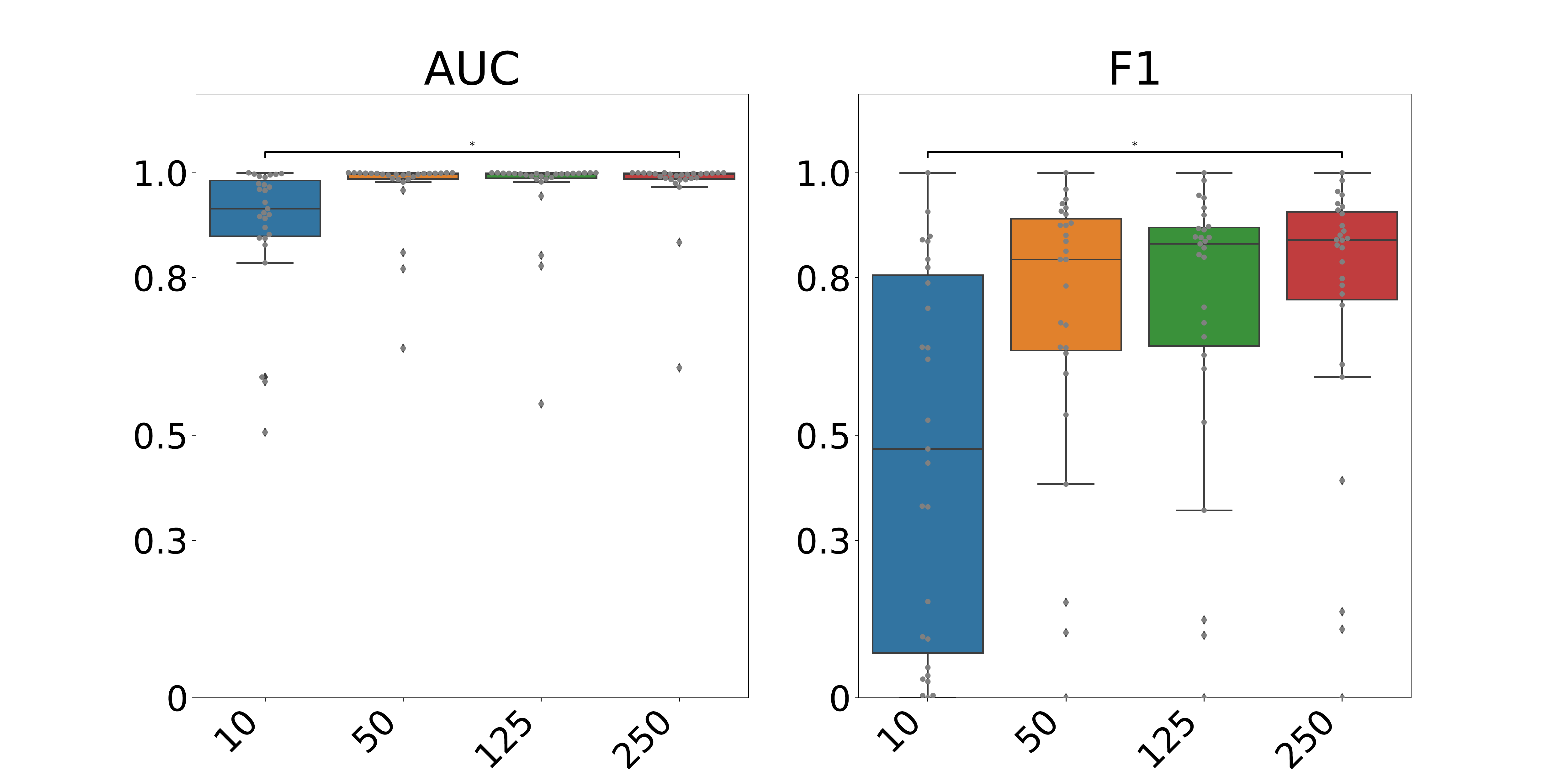}
  \caption{Evaluation of the kernel size in the first convolutional layer of mEEGNet. $\ast$ represents $p<0.05$ in the posthoc test.}
  \label{fig:kernel_size}
\end{figure}

The results of the 5-fold cross-validation with the kernel sizes of 10, 50, 125, and 250 in the temporal direction of the first mEEGNet layer are presented in \wfig{kernel_size}.
The Friedman test showed that the AUC for the 5-fold cross-validation was $\chi^2(3) = 29.1$, $p = 2.17\times 10^{-6}$, and for the F1-value, $\chi^2(3) = 60.2$, $p = 5.29\times 10^{-13}$.
Since significant differences were found in all validations, a post-test was conducted. As shown in \wfig{kernel_size}, a significant difference was found between the kernel sizes of 250 and 10. From the box plots, it was observed that the median F1-value tended to increase as the kernel size increased.
Similar results were obtained in the leave-one-case-out method. In the AUC of the leave-one-case-out validation, $\chi^2(3) = 25.5$, $p=1.21\times 10^{-5}$, and in the F1-value, $\chi^2(3) = 25.6$, $p=1.14\times10^{-5}$. As a result of the post-hoc test, a significant difference was found between the kernel sizes of 250 and 10.

\section{Discussion}

We have assessed evaluation scores (AUC, F1-values, sensitivity and specificity) for two-class classification models with different loss functions. The experimental results showed that the proposed mEEGNet with the binary cross-entropy provided the best performance in the two criteria. Additionally, kernels that were too short (size of 10) led to significant deterioration in the classification performance.

\subsection{Weighted Loss Functions}
\label{s:discussion_wt}

The focal loss and the class-balanced focal (CBFocal) loss were proposed for datasets with imbalanced labels \citep{Lin2017,Cui2019}. \citet{Cui2019} proposed to characterize the dataset by the so-called imbalance factor of a dataset, defined as the number of training samples in the largest class divided by the smallest. The CBFocal loss showed better classification accuracy than the BCE and Focal in an imbalance factor range of 1 to 200. Although our dataset falls into this range--the imbalance factor in terms of time is 36.006 (see Table \ref{tab:dataset})--there was almost no difference in performance between the three types of loss functions compared. Thus, the CBLoss was not as effective as expected. This might be due to the selection of parameters, which should need fine-tunig.

\subsection{Effect of the Kernel Size}
\label{s:discussion_ks}

From Fig.~\ref{fig:kernel_size}, it can be confirmed that the performance of mEEGNet with a kernel size of 10 is significantly lower than those of mEEGNet with a larger kernel size in both 5-fold cross-validation and leave-one-case-out validation. This result may be explained by the frequency bandwidth extractable depending on the kernel size. The dataset used in this paper all consist of EEGs with a sampling frequency of 500 Hz. Therefore, when the kernel size is 250, features with a 2 Hz or higher frequency can be extracted. On the other hand, if the kernel size is 10, only features in the frequency domain above 50 Hz can be extracted. In EEG analysis, the frequency band below 40 Hz is often used \citep{Craik2019} because essential features often exist in such a low-frequency band. This could lead to a significant performance difference between the mEEGNet with a kernel size of 10 in the first layer and those with a larger kernel size.

\subsection{Advantage of mEEGNet in Terms of Number of Parameters}
\label{s:discussion_svm}

\begin{table}[t]
  \centering
  \caption{Number of parameters of CNNs used in other studies for detecting abnormal epileptic EEGs, where the input sample size is 500 and the number of electrodes are 16.}
  \label{tab:parameters}
  \begin{tabular}{l|cr} \hline
  Network & \# of parameters
  \\ \hline
  ScalpNet~\citep{Sakai2020} & $10,012,752$ \\
  Hossain et al.~\citep{Hossain2019} & $192,172$\\
  Zhou et al.~\citep{Zhou2018} & $186,296$ \\ \hline
  mEEGNet (Proposed) & $6,784$ \\ \hline
  \end{tabular}
\end{table}




Results of the AUC, F1-values, and sensitivity tests showed that the proposed mEEGNet achieved the best performance. However, as observed in Tables \ref{tab:5fold} and \ref{tab:loco}, ScalpNet and the Zhou-freq model performed similarly to the mEEGNet. However, we would like to conclude that mEEGNet is the best in overall performance because the number of parameters of mEEGNet is the significantly smallest, as summarized in Table \ref{tab:parameters}. It should be noted that the number of parameters of mEEGNet is about one-third that of Zhou et al.'s model and about one over 1,500 that of ScalpNet.

\subsection{Characteristics of Abnormality Detection}

\begin{figure}[t]
    \centering
    \includegraphics[width=.7\textwidth]{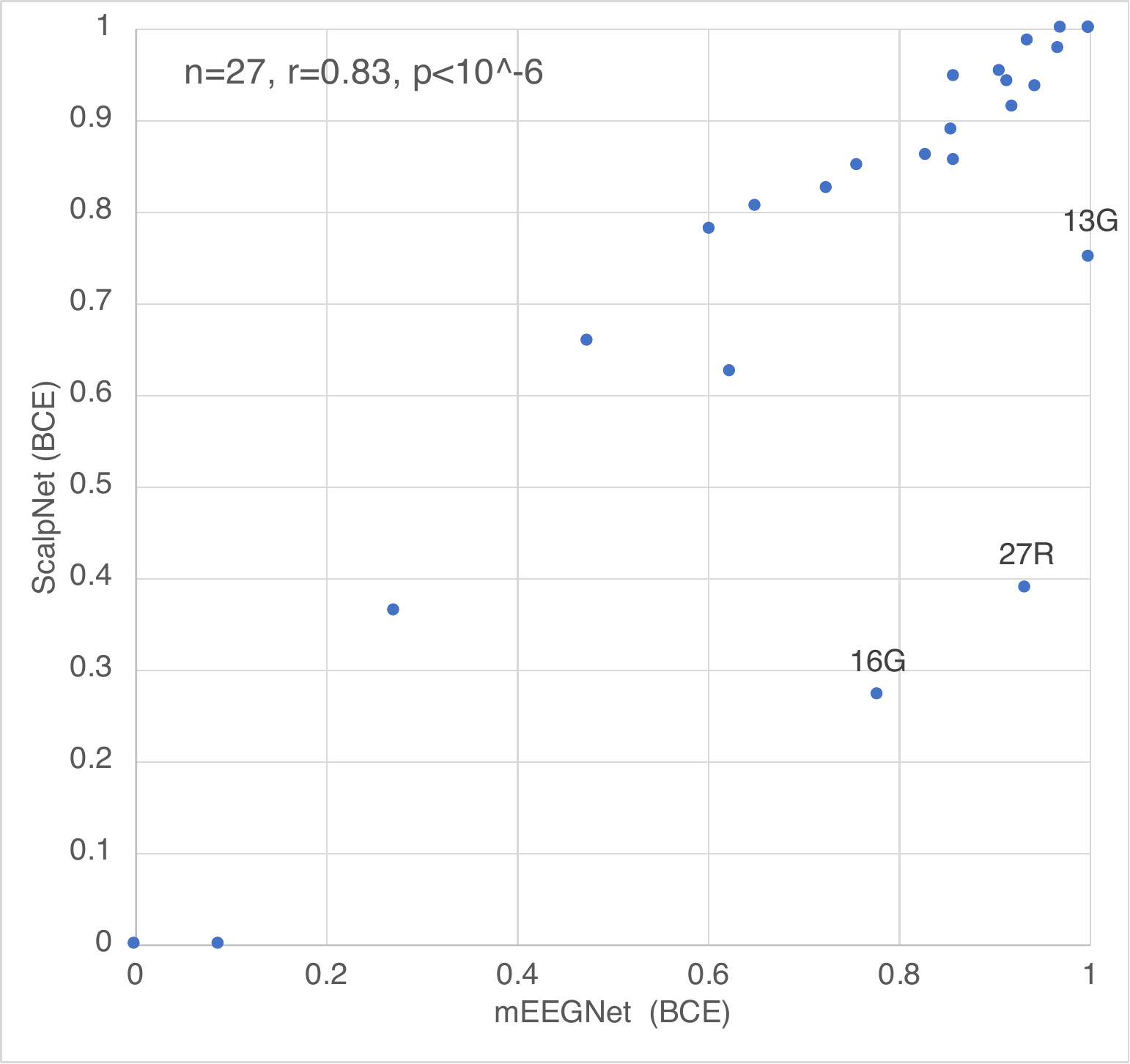}
    \caption{Correlaton of sensitivity}
    \label{fig:correlation}
\end{figure}

It is also essential to discuss the behavior of abnormality detection of the proposed mEEGNet. To evaluate the behavior, we consider some representative cases. We focus on sensitivity because specificity values are nearly one for all models, and sensitivity is an essential measure in the abnormality detection application. Figure \ref{fig:correlation} is the scatter plot of sensitivity values of the mEEGNet (BCE) and ScalpNet (BCE) for each case in the leave-one-case-out validation. It is observed from this figure that the sensitivity is strongly correlated between two models ($r=0.83$, $p<10^{-6}$), whereas three cases (13G, 16G, and 27R) showed a relatively notable difference in the sensitivity.

\begin{figure}[t]
    \centering
    \subfloat[EEG No.~16G]{\includegraphics[width=\textwidth]{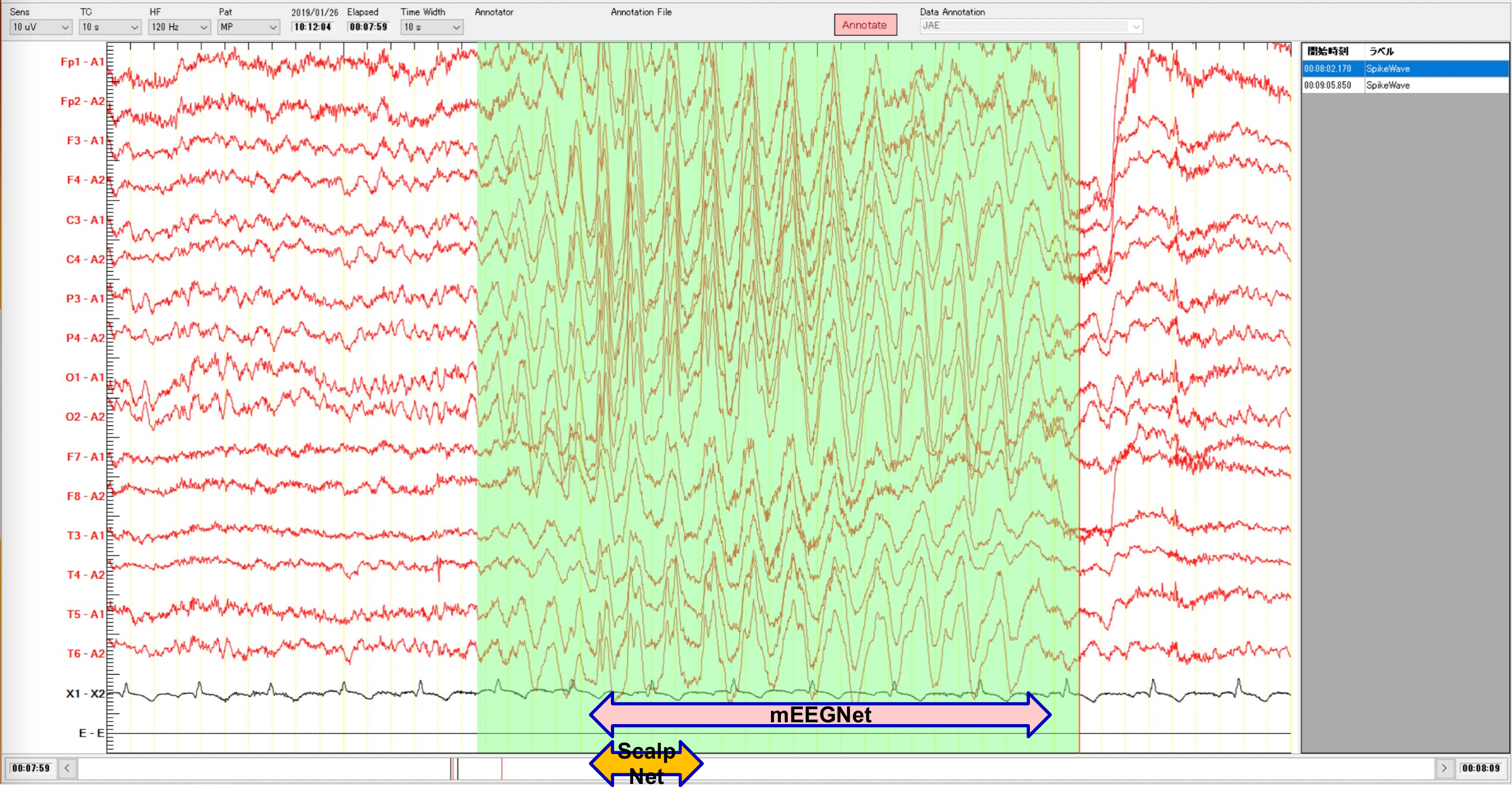}\label{fig:detection:16G}}
    
    \subfloat[EEG No.~27R]{\includegraphics[width=\textwidth]{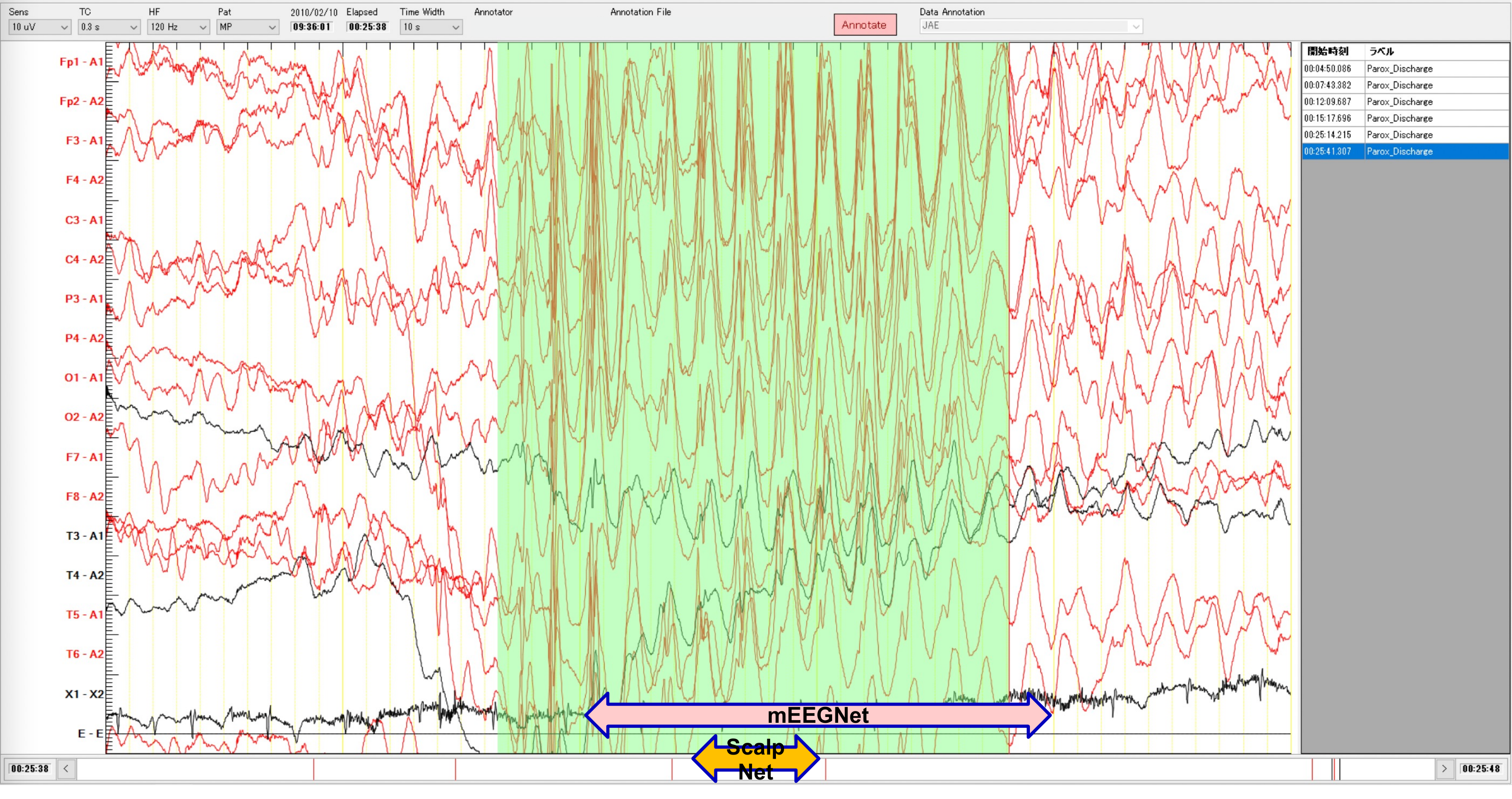}\label{fig:detection:27R}}
    \caption{Examples of detected intervals of abnormalities. The green shadow indicates the intervals labeled by an epileptlogist. The intervals detected by mEEGNet (BCE) and ScalpNet (BCE) are indicated by the double-headed arrows with the model names.}
    \label{fig:detection}
\end{figure}

EEG No.~13G contains one interval ($443.680 \le t \le 448.424$ sec) of abnormality (spike-and-wave). Both mEEGNet (BCE) and ScalpNet (BCE) detected this interval but with different sensitivity. The mEEGNet returned a range of $444 \le t \le 448$ sec as the abnormal interval, whereas the ScalpNet detected $444 \le t \le 447$ sec, which does not cover the labeled interval and is shorter than mEEGNet. For the other two cases (16G and 27R), the mEEGNet's sensitivity for abnormalities is stronger than the ScalpNet. Illustrative examples are provided in Fig.~\ref{fig:detection}. The double-headed arrows indicate the intervals detected by the two models. Figures \ref{fig:detection:16G} and \ref{fig:detection:27R} show that the interval detected by the ScalpNet is much shorter than that by the mEEGNet. A possible reason may be in the first layer of the ScalpNet, which sums $5 \times 4$-images (EEG maps, as illustrated in Fig.~\ref{fig:scalpnet}) over a whole time instances in the analyzing segment. In other words, the first layer is the only layer to implement convolutions in time. On the other hand, mEEGNet includes two layers (Conv2D and SeperableConv2D as shown in Table \ref{tab:mEEGNet}) for time convolutions, which may be more effective in detecting temporal abnormalities in EEGs.

\section{Conclusion}
This paper has proposed using the mEEGNet to detect abnormalities in EEGs of absence epilepsy (CAE and JAE). The structure of the mEEGNet is a variant of the EEGNet, which was designed in the context of brain-computer interfacing. With our clinical dataset, we have confirmed that the proposed variant of EEGNet is also effective in detecting abnormal patterns in EEGs of absence epilepsy patients. Even though the dataset used for training is highly imbalanced, the proposed mEEGNet achieved notable AUC, F1, and sensitivity scores. Furthermore, the number of parameters is significantly smaller than conventional models that have been used for EEG in epilepsy. However, we have only tested the proposed mEEGNet with a limited type of epilepsy: JAE and CAE. EEGs in other kinds of epilepsy should be examined in the near future. Although the mEEGNet was applied to detect abnormal patterns for absence epilepsy, we can conjecture that the EEGNet family might be effective in seizure detection in general and partial epilepsy. Moreover, since abnormal patterns in either JAE or CAE are intermittent and infrequent, the detection problem is similar to anomaly detection. Thus, machine learning techniques for anomaly detection may works for abnormal EEG pattern detection in epilepsy. Further investigations are required.

\section*{Acknowledgements}

The authors thank Toshiki Orihara for creating the EEG screenshots. This work was supported in part by JST CREST (JPMJCR1784).

\bibliography{ref,fukumori}

\end{document}